\documentclass[a4paper]{jpconf}

\usepackage{graphicx}

\def\nuc#1#2{${}^{#1}$#2}
\def\mee{$\langle m_{\beta\beta} \rangle$}

\def\BBz{$\beta\beta(0\nu)$}

\def\BB{$\beta\beta$}

\def\Tz{$T^{0\nu}_{1/2}$}

\def\ms{$\delta m_{\rm sol}^{2}$}

\def\qval{$Q_{\beta\beta}$}                 % The Q-value
\def\be{\begin{equation}}
\def\ee{\end{equation}}
\def\cpKkgy{cnts/(keV kg y)}

\def\cpRty{cnts/(ROI t y)}
\def\onecpRty{1~cnt/(ROI t y)}

\def\ppc{P-PC}                          % P-type Point Contact
\def\MJ{{\sc Majorana}}             %Majorana project name
\def\DEM{{\sc Demonstrator}}             %Demonstrator in small caps

\begin{document}
\title{Initial Results from the \MJ\ \DEM}

\author{S.R.~Elliott$^{1}$,  N.~Abgrall$^{2}$,  I.J.~Arnquist$^{3}$, F.T.~Avignone~III$^{4,5}$, A.S.~Barabash$^{6}$, F.E.~Bertrand$^{5}$, A.W.~Bradley$^{2}$, V.~Brudanin$^{7}$, M.~Busch$^{8,9}$, M.~Buuck$^{10}$, T.S.~Caldwell$^{11,9}$, Y-D.~Chan$^{2}$, C.D.~Christofferson$^{12}$, P.-H.~Chu$^{1}$, C. Cuesta$^{10}$, J.A.~Detwiler$^{10}$, C. Dunagan$^{12}$, Yu.~Efremenko$^{13}$, H.~Ejiri$^{14}$, A. Fullmer$^{15,9}$, A.~Galindo-Uribarri$^{5}$, T.~Gilliss$^{11,9}$, G.K.~Giovanetti$^{16}$, M.P.~Green$^{15,9,5}$, J. Gruszko$^{10}$, I.S.~Guinn$^{10}$, V.E.~Guiseppe$^{4}$, R.~Henning$^{11,9}$, E.W.~Hoppe$^{3}$, M.A.~Howe$^{11,9}$, B.R.~Jasinski$^{17}$, K.J.~Keeter$^{18}$, M.F.~Kidd$^{19}$, S.I.~Konovalov$^{6}$, R.T.~Kouzes$^{3}$, J. Leon$^{10}$, A.M.~Lopez$^{13}$, J.~MacMullin$^{11,9}$, R.D.~Martin$^{20}$, R. Massarczyk$^{1}$, S.J.~Meijer$^{11,9}$, S.~Mertens$^{2,21}$, J.L.~Orrell$^{3}$, C. O'Shaughnessy$^{15,9}$, A.W.P.~Poon$^{2}$, D.C.~Radford$^{5}$, J.~Rager$^{11,9}$, K.~Rielage$^{1}$, R.G.H.~Robertson$^{10}$, E. Romero-Romero$^{13,5}$, B.~Shanks$^{11,9}$, M.~Shirchenko$^{7}$, A.M.~Suriano$^{12}$, D.~Tedeschi$^{4}$, J.E.~Trimble$^{11,9}$, R.L.~Varner$^{5}$, S. Vasilyev$^{7}$, K.~Vetter$^{2,21}$, K.~Vorren$^{11,9}$, B.R.~White$^{1}$, J.F.~Wilkerson$^{11,9,5}$, C. Wiseman$^{4}$, W.~Xu$^{11,9,22}$, E.~Yakushev$^{7}$, C.-H.~Yu$^{5}$, V.~Yumatov$^{6}$, I.~Zhitnikov$^{7}$}

\address{$^{1}$~Los Alamos National Laboratory,  Los Alamos,  NM,  USA}
\address{$^{2}$~Nuclear Science Division,  Lawrence Berkeley National Laboratory,  Berkeley,  CA,  USA}
\address{$^{3}$~Pacific Northwest National Laboratory,  Richland,  WA,  USA}
\address{$^{4}$~Department of Physics and Astronomy,  University of South Carolina,  Columbia,  SC,  USA}
\address{$^{5}$~Oak Ridge National Laboratory,  Oak Ridge,  TN,  USA}
\address{$^{6}$~National Research Center ``Kurchatov Institute'' Institute for Theoretical and Experimental Physics, Moscow, Russia}
\address{$^{7}$~Joint Institute for Nuclear Research,  Dubna,  Russia}
\address{$^{8}$~Department of Physics,  Duke University,  Durham,  NC,  USA}
\address{$^{9}$~Triangle Universities Nuclear Laboratory,  Durham,  NC,  USA}
\address{$^{10}$~Center for Experimental Nuclear Physics and Particle Astrophysics,  and Department of Physics,  University of Washington,  Seattle,  WA,  USA}
\address{$^{11}$~Department of Physics and Astronomy,  University of North Carolina,  Chapel Hill,  NC,  USA}
\address{$^{12}$~South Dakota School of Mines and Technology,  Rapid City,  SD,  USA}
\address{$^{13}$~Department of Physics and Astronomy,  University of Tennessee,  Knoxville,  TN,  USA}
\address{$^{14}$~Research Center for Nuclear Physics, Osaka University, Ibaraki, Osaka, Japan}
\address{$^{15}$~Department of Physics,  North Carolina State University,  Raleigh,  NC,  USA}
\address{$^{16}$~Department of Physics, Princeton University, Princeton, NJ, USA}
\address{$^{17}$~Department of Physics,  University of South Dakota,  Vermillion,  SD,  USA}
\address{$^{18}$~Department of Physics,  Black Hills State University,  Spearfish,  SD,  USA}
\address{$^{19}$~Tennessee Tech University, Cookeville, TN, USA}
\address{$^{20}$~Department of Physics, Engineering Physics and Astronomy, Queen's University, Kingston, ON, Canada}
\address{$^{21}$~Permanent address: Max-Planck-Institut f\"{u}r Physik, and Physik Department and Excellence Cluster Universe, Technische Universit\"{a}t, M\"{u}nchen, Germany}
\address{$^{22}$~Permanent address: Department of Physics,  University of South Dakota,  Vermillion,  SD,  USA}
\address{$^{23}$~Alternate address:  Department of Nuclear Engineering,  University of California,  Berkeley,  CA,  USA}

\ead{elliotts@lanl.gov}

\begin{abstract}
Neutrinoless double-beta decay searches seek to determine the nature of neutrinos, the existence of a lepton violating process, and the effective Majorana neutrino mass. The {\sc Majorana} Collaboration is assembling an array of high purity Ge detectors to search for neutrinoless double-beta decay in $^{76}$Ge. The {\sc Majorana Demonstrator} is composed of 44.8~kg (29.7 kg enriched in $^{76}$Ge) of Ge detectors in total, split between two modules contained in a low background shield at the Sanford Underground Research Facility in Lead, South Dakota. The initial goals of the {\sc Demonstrator} are to establish the required background and scalability of a Ge-based, next-generation, tonne-scale experiment. Following a commissioning run that began in 2015, the first detector module started physics data production in early 2016. We will discuss initial results of the Module 1 commissioning and first physics run, as well as the status and potential physics reach of the full {\sc Majorana Demonstrator} experiment. The collaboration plans to complete the assembly of the second detector module by mid-2016 to begin full data production with the entire array.

\end{abstract}

\section{Introduction}
Neutrinoless double-beta (\BBz) decay searches represent the only viable experimental method for testing the Majorana nature of the neutrino~\cite{Zralek1997}. The observation of this process would immediately imply that lepton number is violated and that neutrinos are Majorana particles~\cite{sch82}. A measurement of the \BBz\ decay rate may also yield information on the absolute neutrino mass. Measurements of atmospheric, solar, and reactor neutrino oscillation~\cite{Beringer2012} indicate a large parameter space for the discovery of \BBz\ decay just beyond the current experimental bounds below \mee\ $\sim$50 meV. Moreover, evidence from the SNO experiment~\cite{Ahm04} of a clear departure from non-maximal mixing in solar neutrino oscillation implies a minimum effective Majorana neutrino mass of $\sim$15 meV for the inverted mass ordering scenario. This target is within reach of next-generation \BBz\ searches. An experiment capable of observing this minimum rate would therefore help elucidate the Majorana or Dirac nature of the neutrino for inverted-hierarchical neutrino masses. Even for the normal hierarchy, these experiments would improve the existing sensitivity by $\approx$1 order of magnitude. A nearly background-free tonne-scale \nuc{76}{Ge} experiment would be sensitive to effective Majorana neutrino masses below $\sim$15~meV. For recent comprehensive experimental and theoretical reviews, see Refs.~\cite{avi08, bar11, Rode11, Elliott2012, Vergados2012, Cremonesi2013, Schwingenheuer2013,Elliott2015}.

Recent developments in germanium detector technology make a large-scale \BBz\ decay search feasible using \nuc{76}{Ge}. In these proceedings we describe the \MJ\ \DEM, an experimental effort completing construction during 2016, whose goal is to demonstrate the techniques required for a next-generation \BBz\ decay experiment with enriched Ge detectors.  A complementary \nuc{76}{Ge} effort, the GERDA experiment~\cite{Ackermann2013,Agostini2016}, is presently operating in the Laboratori Nazionali del Gran Sasso (LNGS).  

\section{Experimental Overview}
The \MJ\ \DEM\ is an array of isotopically enriched and natural Ge detectors that will search for the decay of isotope \nuc{76}{Ge}~\cite{Abgrall2014}. The primary goal of the \DEM\ is to demonstrate a path forward to achieving a background rate low enough (at or below \onecpRty\ in the 4 keV region of interest (ROI) around the 2039-keV Q-value for \nuc{76}{Ge}) to ensure the feasibility of a future Ge-based \BB\ experiment to probe the inverted-hierarchy parameter space of neutrino mass. 

\MJ\ utilizes the demonstrated benefits of enriched high-purity germanium (HPGe) detectors. These include intrinsically low-background source material, understood enrichment chemistry, excellent energy resolution, and sophisticated event reconstruction.  We have assembled a modular instrument composed of two cryostats built from ultra-pure electroformed copper, with each cryostat housing over 20 kg of P-type, point-contact (\ppc) HPGe detectors~\cite{luk89,Barbeau2007,Aguayo2011}.

 \ppc\ detectors were chosen after extensive R\&D and each has a mass of about 0.6-1.0 kg. The two cryostats contain 35 detectors with a total mass of 29.7 kg, fabricated with Ge material enriched to 88\% in isotope 76, and 15.1 kg fabricated from natural Ge (7.8\% \nuc{76}{Ge}).  The 74.5\% yield of converting initial material into Ge diodes is the highest achieved to date.

\begin{figure}[h]
 \centering
 \includegraphics[width=0.65\columnwidth, keepaspectratio=true]{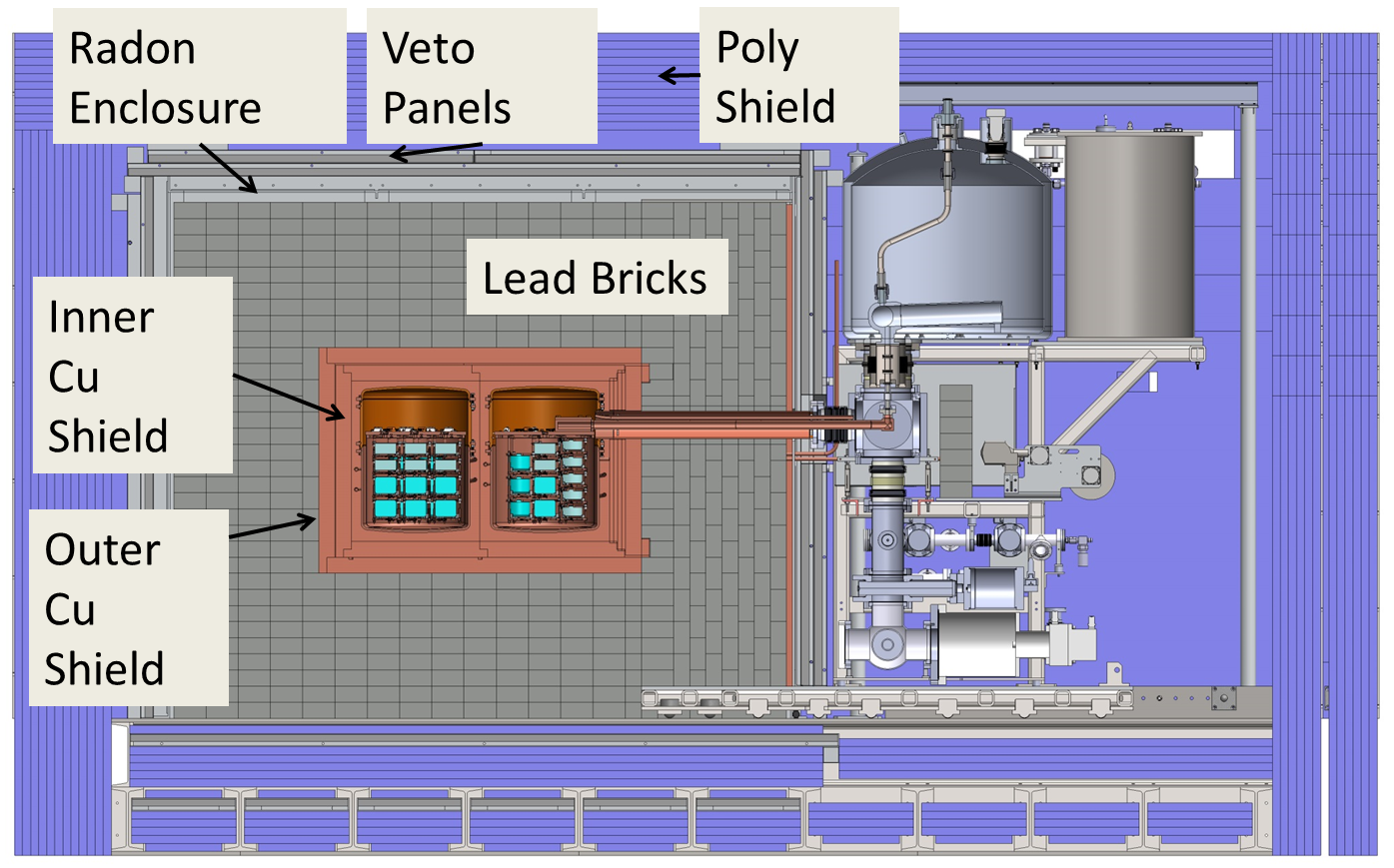}
 \includegraphics[width=0.3\columnwidth, keepaspectratio=true]{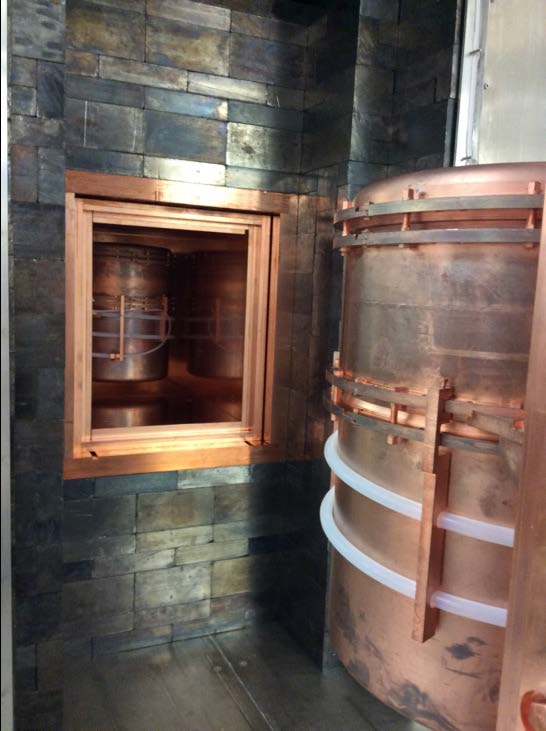}
 \caption{Left: The \DEM\ shield system in cross section, shown with both cryostats installed. Right: A photo of Cryostat 1 installed within the shield and Cryostat 2 just before insertion.}
 \label{Fig:layout}
\end{figure}

The experimental layout is shown in Fig.~\ref{Fig:layout}. Starting from the innermost cavity, the cryostats are surrounded by an inner layer of electroformed copper, an outer layer of commercially obtained Oxygen-Free High thermal Conductivity (OFHC) copper, high-purity lead, an active muon veto, borated polyethylene, and polyethylene. The cryostats, copper, and lead shielding are all enclosed in a radon exclusion box that is purged with liquid nitrogen boil-off. The experiment is located in a clean room at the 4850-foot level (1478 m) of the Sanford Underground Research Facility (SURF) in Lead, South Dakota~\cite{Heise2015}.

\section{The Data Sets and Analysis}
As a first step, a prototype module was constructed using a cryostat made of commercial copper. It was loaded with three strings of natural-isotopic-abundance HPGe and placed into the shielding. Data was collected with this module from June 2014 through July 2015. It served as a test bench for mechanical designs, fabrication methods, and assembly procedures for the construction of the electroformed-copper Modules~1~\&~2. In addition, the prototype also tested DAQ, data production and analysis tools.

Following the prototype run, construction began on two modules with electroformed-copper cryostats.  The first, Module~1, was assembled in 2015. Module~1 houses 16.8~kg of enriched germanium detectors and 5.7~kg of natural germanium detectors. Module~1 was moved into the shield, and data taking began during 2015. Module~2 supports 12.8~kg of enriched and 9.4~kg of natural Ge detectors, and has been assembled and began taking data in August 2016. 

The data from Module 1 are divided into two datasets: dataset~0 (DS0) and dataset~1 (DS1). DS0 was a set of commissioning runs used to test analysis and data production and to get a first look at backgrounds, and corresponds to data taken from June 26, 2015 to October 7, 2015. In Fall 2015, we implemented the following planned improvements:
\begin{itemize}
\item Installed the electroformed inner copper shield. Built with the final pieces of electroformed copper, the inner shield was not yet ready during the initial construction of Module~1.
\item Added shielding within the vacuum of the cryostat cross arm.
\item Replaced the cryostat Kalrez seal with PTFE, which has much better radiopurity and much lower mass. This is expected to result in a 3 orders of magnitude background reduction in the ROI.
\item Repaired non-operating channels.
\end{itemize}

These changes define the difference between the data sets, with DS1 being the lower background configuration of the two. Hence, DS1 is the dataset that is being used to determine the background. DS1 described here was taken from December 31, 2015 to April 14, 2016. After April 14, data taking continued with a newly implemented blinding scheme. On May 24, 2016 we implemented improved waveform recording intended to improve the removal of surface $\alpha$ events (discussed below and in Ref.~\cite{Gruszko2016}). With that change, DS1 was considered complete. Module~1 operation was temporarily ceased on July 14, 2016 for the installation of Module~2. After both modules were installed, data collection resumed in late August.

Data from the \DEM\ are first filtered to remove non-physical waveforms. Next we remove events whose waveforms are typical of multi-site energy deposits. Double-beta decay events are characterized as single-site events because the range of electrons is small compared to that of a typical Compton-scattering background gamma. Using pulse shape discrimination methods~\cite{dusa09,Cuesta2016}, and vetoing coincidences between two or more detectors, it is possible to reject $>$90\% of multi-site events while retaining 90\% of single-site events and reducing the Compton continuum at \qval\ by $>$50\% in the case of backgrounds from the $^{228}$Th calibration source.

Finally we remove events that arise from $\alpha$ particles impinging upon the passivated surface of the detector. For such events, some of the charge drift takes place within the detector bulk, but much also takes place along the detector's surface with a significantly different drift speed. The reconstructed energy of these events corresponds to the fraction of energy collected within the shaping time of our energy filter, resulting in energy degradation that sometimes populates the critical region of interest. However, the waveform distortion resulting from the slow surface drift mobility of the rest of the drifting charge permits an extremely effective reduction of this potential background using pulse shape discrimination~\cite{Gruszko2016}. 

We are continuing to improve the waveform analysis for both of these pulse-shape-based cuts. The background model arising from our radioassay program~\cite{Abgrall2016} and our analysis cuts is summarized in Ref.~\cite{Cuesta2016a}.

\section{Discussion of Early Results}
\begin{figure}[h]
 \centering
 \includegraphics[width=0.5\columnwidth, keepaspectratio=true]{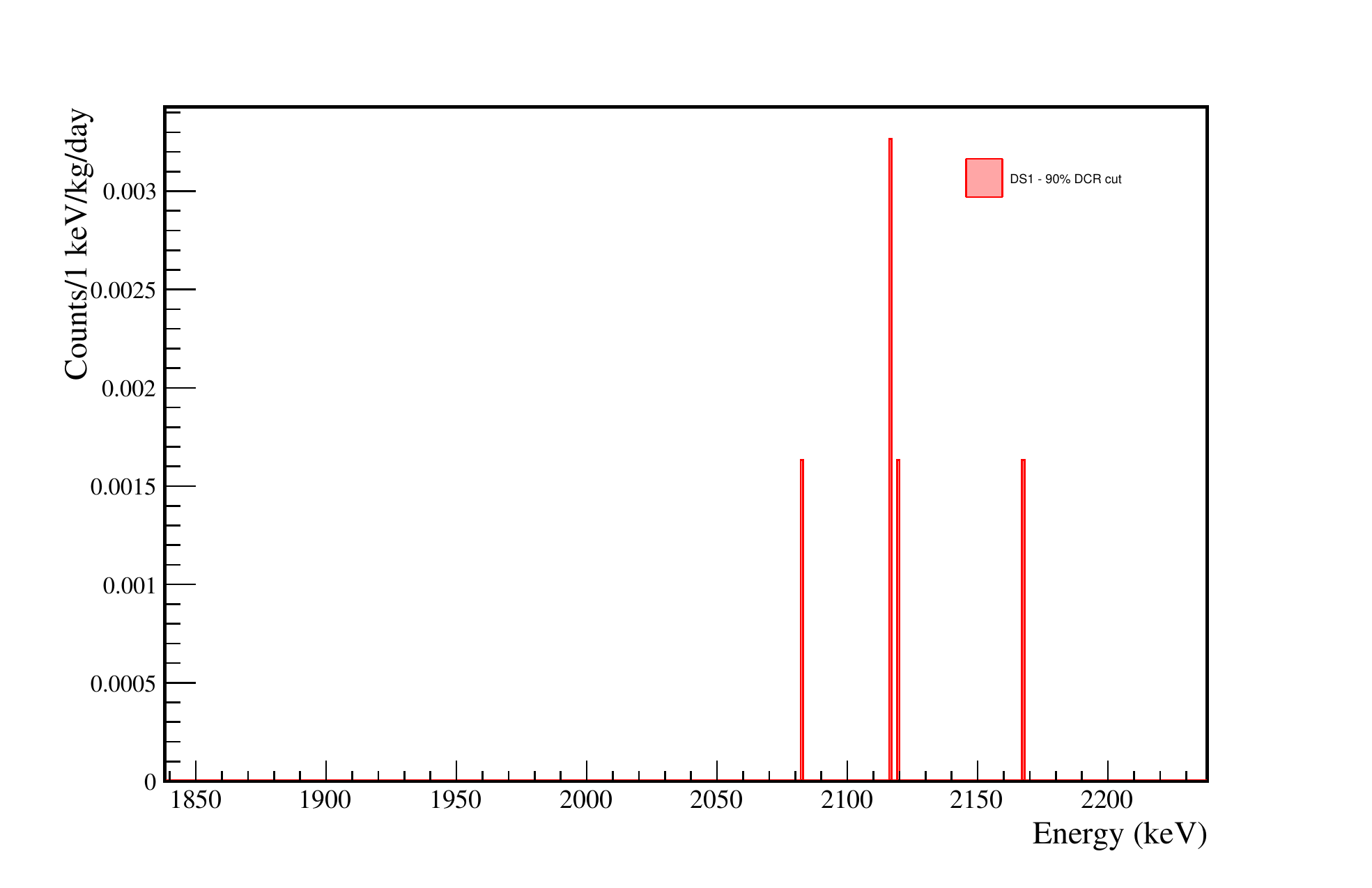}
 \includegraphics[width=0.45\columnwidth, keepaspectratio=true]{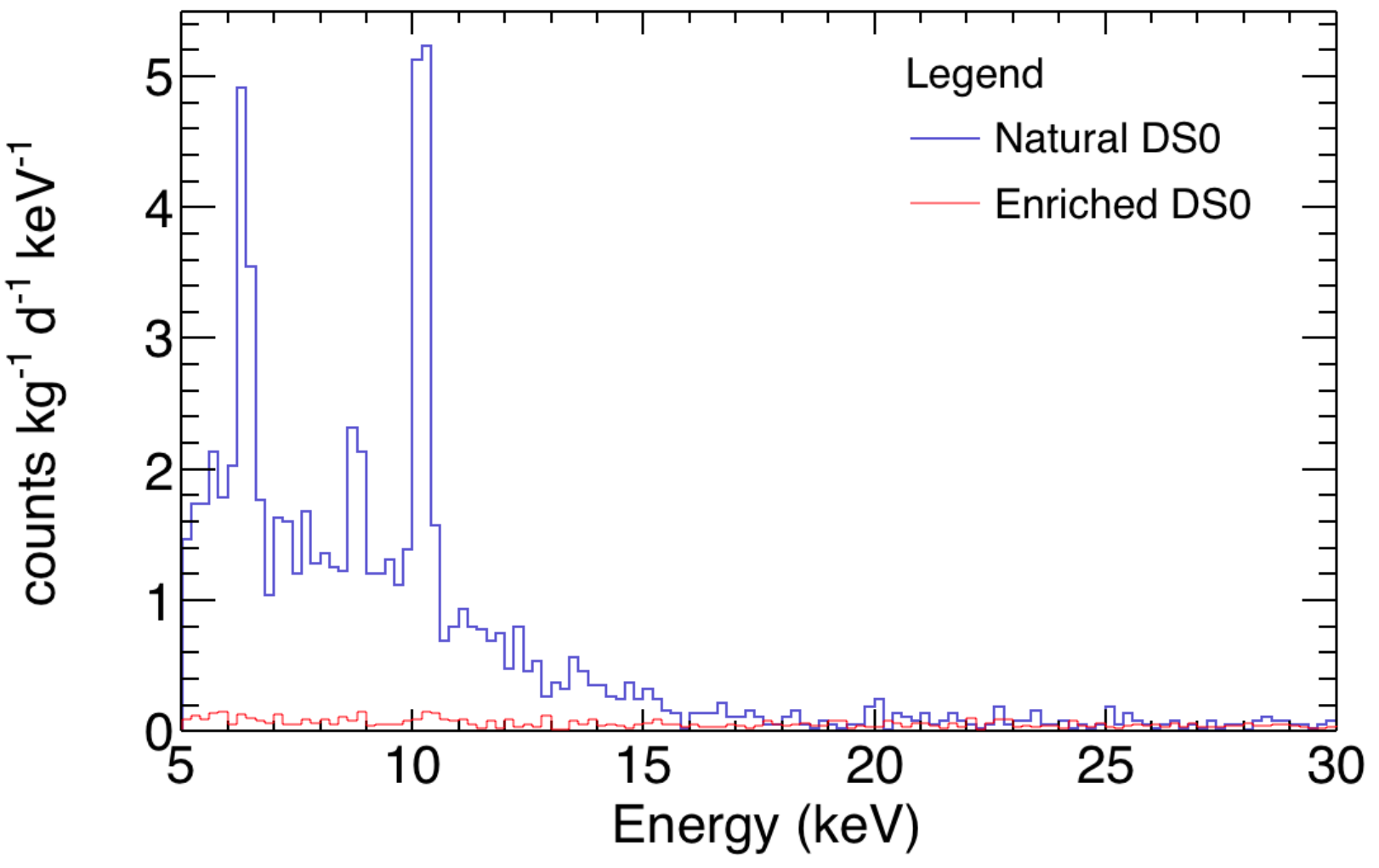}
 \caption{Left: The spectrum between 1838 and 2238 keV after all cuts from DS1. Right:  Low energy spectra from 192 kg d of natural (blue) and 478 kg d of enriched (red) detector data from DS0. Cosmogenic isotopes in the natural detectors produce the 10.3 keV $^{68}$Ge, the 8.9~keV $^{65}$Zn, and the 6.5 keV $^{55}$Fe X-ray peaks along with the tritium beta decay continuum. The FWHM of the 10.3 keV peak is $\sim$0.25 keV. The spectrum shown does not include an efficiency correction.}
 \label{Fig:Background}
\end{figure}

After all cuts in DS1, there are 5 events within a 400 keV window centered on \qval. (See Fig.~\ref{Fig:Background}.) This results in a background index of $(7.5^{+4.5}_{-3.4}\times10^{-3}$ \cpKkgy). In the context of background rates for a next-generation \BBz\ experiment, this level corresponds to $23^{+13}_{-10}$ \cpRty\ at 68\% CL for a 3.1 keV ROI. 

The exposures are 1.37 kg-y and 1.66 kg-y for DS0 and DS1 respectively. For DS1, the results here only reflect data collected up through April 14, 2016. The efficiency for \BBz\ is $0.61\pm0.04$ resulting from cuts due to resolution (0.84), the probability that \BB\ results in a full energy deposit (0.90), the single-site waveform cut (0.90), and the surface-$\alpha$ cut (0.90). Although the background is lower in DS1, it is still low enough in DS0 that we can combine the two to derive a half-life limit. As there are no candidate events within the 3.1-keV region of interest near \qval, we derive a lower limit on the \Tz\ of $3.7\times10^{24}$ y with  90\% CL. This analysis was based on an open data set. The expected background within the region of interest is still low enough that \Tz\ will increase nearly linearly with exposure. With nearly 30 kg of \nuc{76}{Ge}, we project sensitivity near $10^{26}$ y with 3-5 years of data.

The low-energy spectrum from DS0 is shown in Fig.~\ref{Fig:Background}. A similar low-energy analysis for DS1 is in progress. The data cuts are similar to those described above, except that an additional cut associated with {\em slow pulses} is implemented. Slow pulse waveforms with rise-times of $\mathcal{O}$(1$\mu$s) or longer constitute a significant background below 30 keV, as recognized by previous low-energy PPC Ge-detector experiments \cite{Aalseth2014, Aguayo2013,Giovanetti2015}. Slow pulses are energy degraded events which originate in low-field detector regions near the surface dead layer where diffusion is the dominant mode of charge transport. At energies $<$10 keV, discriminating slow pulses using pulse rise-time estimators becomes increasingly difficult due to the worsening signal to noise ratio. A more robust parameter, $T/$E, is obtained by finding the maximum value of a triangle filter ($T$) and normalizing the result by the energy ($E$), determined offline using a trapezoidal filter. This parameter exhibited good separation down to $\sim$3 keV, at which point $T$ is more dependent on noise than rise-time. 

The spectrum from the enriched detectors shows a significant reduction in the contribution due to cosmogenic isotopes owing to our efforts to minimize cosmic ray exposure of the enriched material. These data are being used to derive limits on several hypothetical physics processes including bosonic dark matter interactions, axions of solar origin, Pauli Exclusion Principle violating electronic decays, and electron decay.

\section{Summary}
The goal of the \MJ\ \DEM\ is to show that backgrounds can be reduced to a value low enough to justify a large \BBz\ experiment using \nuc{76}{Ge}. We have built two modules from ultra-low-background components that contain Ge detectors. The first module has been operating since June 2015 and initial results indicate that at the experiment's start, the background level is very low. We anticipate further reductions by improving our pulse shape analysis algorithms. Furthermore, the background data show a small peak at 2614 keV (\nuc{208}{Tl}). Although the statistics are poor, the area of this peak is such that the ROI background level remaining after the surface-$\alpha$ cut is not inconsistent with being dominated by Compton scattering of the 2614-keV $\gamma$. Efforts to localize, and hopefully remove, the source of these $\gamma$s have begun.The \DEM\ goal is to reach a background of 3 \cpRty\ and as of this writing we have achieved 23 \cpRty. At the present level of background, the limit on \Tz\ is increasing nearly linearly and is projected to reach an asymptotic limit near $10^{26}$ y.

\section*{Acknowledgements}
This material is based upon work supported by the U.S. Department of Energy,  Office of Science,  Office of Nuclear Physics under Award  Numbers DE-AC02-05CH11231,  DE-AC52-06NA25396,  DE-FG02-97ER41041,  DE-FG02-97ER41033,  DE-FG02-97ER41042,  DE-SC0012612,  DE-FG02-10ER41715,  DE-SC0010254,  and DE-FG02-97ER41020. We acknowledge support from the Particle Astrophysics Program and Nuclear Physics Program of the National Science Foundation through grant numbers PHY-0919270,  PHY-1003940,  0855314,  PHY-1202950,  MRI 0923142 and 1003399. We acknowledge support from the Russian Foundation for Basic Research,  grant No. 15-02-02919. We  acknowledge the support of the U.S. Department of Energy through the LANL/LDRD Program. This research used resources of the Oak Ridge Leadership Computing Facility,  which is a DOE Office of Science User Facility supported under Contract DE-AC05-00OR22725. This research used resources of the National Energy Research Scientific Computing Center,  a DOE Office of Science User Facility supported under Contract No. DE-AC02-05CH11231. We thank our hosts and colleagues at the Sanford Underground Research Facility for their support.

\section*{References}
\bibliographystyle{iopart-num}
\bibliography{DoubleBetaDecay.bbl}

\end{document}